\documentclass[prl,aps,reprint,superscriptaddress]{revtex4-1}

\usepackage{graphicx}
\usepackage{dcolumn}
\usepackage{bm}
\usepackage[english]{babel}
\usepackage{color}
\usepackage{amsmath}
\usepackage{amssymb}
\usepackage{graphpap}
\usepackage{units}

\begin{document}

\newcommand{\bise}{Bi$_2$Se$_3$}
\newcommand{\bite}{Bi$_2$Te$_3$}
\newcommand{\sbte}{Sb$_2$Te$_3$}
\newcommand{\bsps}{Bi$_x$Sb$_y$Pb$_z$Se$_3$}
\newcommand{\sro}{ SrTiO$_3$}
\newcommand{\bts}{ Bi$_2$Te$_2$Se}
\newcommand{\bsts}{ Bi$_{2-x}$Sb$_x$Te$_3$Se$_y$}
\newcommand{\rb}[1]{\textcolor{red}{#1}}
\newcommand{\tstern}{\ensuremath{T_2^*} }
\newcommand{\degree}{\ensuremath{^\circ}}
\newcommand*{\fg}[1]{Fig.\thinspace\ref{#1}}
\newcommand*{\fgs}[1]{Figs.\thinspace\ref{#1}}
\newcommand*{\eq}[1]{Eq.\thinspace\ref{#1}}
\newcommand*{\tb}[1]{Tab.\thinspace\ref{#1}}
\newcommand{\todo}[1]{ \textbf{\textcolor{red}{{#1}}}}

%

\title{Spin lifetimes exceeding 12 nanoseconds in graphene non-local spin valve devices}

%

\author{Marc Dr\"{o}geler}\thanks {These authors contributed equally to this work}\affiliation{2nd Institute of Physics and JARA-FIT, RWTH Aachen University, 52074 Aachen, Germany}

\author{Christopher Franzen}\thanks {These authors contributed equally to this work}\affiliation{2nd Institute of Physics and JARA-FIT, RWTH Aachen University, 52074 Aachen, Germany}

\author{Frank Volmer}\affiliation{2nd Institute of Physics and JARA-FIT, RWTH Aachen University, 52074 Aachen, Germany}
\author{Tobias Pohlmann}\affiliation{2nd Institute of Physics and JARA-FIT, RWTH Aachen University, 52074 Aachen, Germany}
\author{Luca Banszerus}\affiliation{2nd Institute of Physics and JARA-FIT, RWTH Aachen University, 52074 Aachen, Germany}
\author{Maik Wolter}\affiliation{2nd Institute of Physics and JARA-FIT, RWTH Aachen University, 52074 Aachen, Germany}
\author{Kenji Watanabe}\affiliation{National Institute for Materials Science, 1-1 Namiki, Tsukuba, 305-0044, Japan}
\author{Takashi Taniguchi}\affiliation{National Institute for Materials Science, 1-1 Namiki, Tsukuba, 305-0044, Japan}
\author{Christoph Stampfer}\affiliation{2nd Institute of Physics and JARA-FIT, RWTH Aachen University, 52074 Aachen, Germany}\affiliation{Peter Gr\"{u}nberg Institute (PGI-9), Forschungszentrum J\"{u}lich, 52425 J\"{u}lich, Germany}
\author{Bernd Beschoten}\affiliation{2nd Institute of Physics and JARA-FIT, RWTH Aachen University, 52074 Aachen, Germany}
\thanks{E-mail address: bernd.beschoten@physik.rwth-aachen.de}

\date{\today}

%

\begin{abstract}
We show spin lifetimes of $\unit[12.6]{ns}$ and spin diffusion lengths as long as $\unit[30.5]{\mu m}$ in single layer graphene non-local spin transport devices at room temperature. This is accomplished by the fabrication of Co/MgO-electrodes on a Si/SiO$_2$ substrate and the subsequent dry transfer of a graphene-hBN-stack on top of this electrode structure where a large hBN flake is needed in order to diminish the ingress of solvents along the hBN-to-substrate interface. Interestingly, long spin lifetimes are observed despite the fact that both conductive scanning force microscopy and contact resistance measurements reveal the existence of conducting pinholes throughout the MgO spin injection/detection barriers. The observed enhancement of the spin lifetime in single layer graphene by a factor of 6 compared to previous devices exceeds current models of contact-induced spin relaxation which paves the way towards probing intrinsic spin properties of graphene.
\end{abstract}
%
\maketitle
%
%
Long electron spin lifetimes as well as spin diffusion lengths are important prerequisites for enabling advanced spintronic devices.\cite{RevModPhys.76.323,ISI000249789600001} Theoretically, graphene should fulfill these requirements thanks to its high charge carrier mobilities and its small spin orbit coupling.\cite{GrapheneSpintronics} Accordingly, initial calculations predicted spin lifetimes of up to $\unit[1]{\mu s}$ for pristine graphene flakes.\cite{PhysRevB.80.041405,PhysRevB.82.085304} However, most electrical spin precession experiments measure spin lifetimes shorter than $\unit[1]{ns}$ and spin diffusion lengths smaller than $\unit[10]{\mu m}$.\cite{Tombros2007,PhysRevB.80.214427,han222109,PhysRevLett.104.187201,doi:10.1021/nl301567n,PhysRevB.87.075455,Avsar2011,PhysRevLett.107.047206,PhysRevB.84.075453,abel:03D115,PhysRevB.87.081405,Neumann2013,APEX.6.073001,Kamalakar2014,1.4893578,1882-0786-7-8-085101,Yamaguchi2012849,doi:10.1021/acsnano.5b02795} Therefore, there has been a strong effort to match spin lifetimes from theory to experimental values. Along this line, theoretical studies propose novel spin scattering mechanisms such as resonant spin scattering by magnetic impurities \cite{PhysRevLett.112.116602} or entanglement between spin and pseudospin by random spin orbit coupling,\cite{Tuan2014} which yield calculated spin lifetimes in the experimentally observed range. On the other hand, several experimental and theoretical studies demonstrate that the measured spin lifetimes are not intrinsic to graphene but are rather limited by invasive contacts.\cite{PhysRevLett.105.167202,PhysRevB.88.161405,PhysRevB.90.165403,PhysRevB.86.235408,PhysRevB.91.241407,PhysRevB.92.201410,2015arXiv151202255A}

One way to diminish the effect of invasive contacts is the increase of the transport channel length.\cite{PhysRevB.86.235408} For example, at a separation of $\unit[16]{\mu m}$ between injection and detection electrodes the spin lifetime was slightly pushed over the $\unit[1]{ns}$-benchmark in graphene grown by chemical vapor deposition.\cite{Kamalakar2015} If the graphene flake in such long-distance devices is additionally encapsulated by hexagonal boron nitride (hBN), spin lifetimes of $\unit[2]{ns}$ and spin diffusion lengths exceeding $\unit[12]{\mu m}$ are measured at room temperature.\cite{PhysRevLett.113.086602} Another approach for less invasive contacts is to avoid direct deposition of electrode materials on top of the graphene flake. We therefore introduced a bottom-up fabrication technique in which Co/MgO-electrodes are fabricated on a substrate and thereafter a PMMA-hBN-graphene stack is transferred on top of this electrode structure. This approach is advantageous as e-beam lithography, wet-chemical processing, and unfavorable growing mechanisms of many materials on top of graphene can significantly reduce the quality of the graphene-to-electrode-interface.\cite{Volmer2015Synthetic} With this procedure we are able to reproducibly build devices showing nanosecond spin lifetimes at room temperature even in case of short transport channel lengths of $\unit[2-3.5]{\mu m}$.\cite{Droegeler2014,PSSB201552418}

\begin{figure}[t]
	\includegraphics*[width=\linewidth]{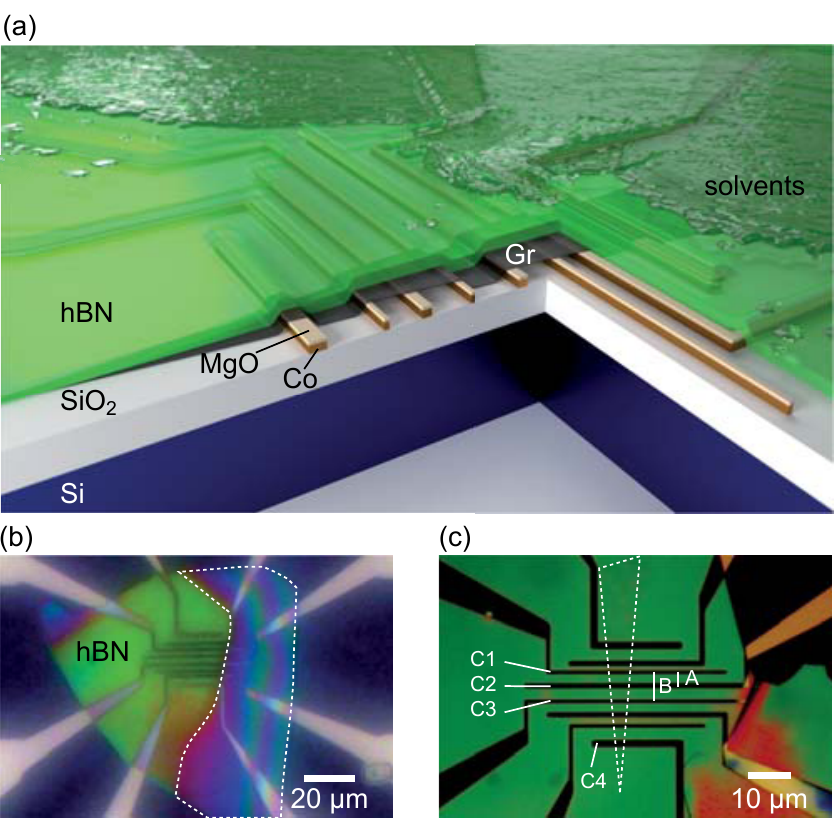}
	\caption{(Color online) (a) Schematic illustration of the spin transport device. When using large hBN flakes the graphene (Gr) is well protected against solvents during the release of the supporting membrane. (b) Optical image of a finished device while the PMMA membrane is dissolved in acetone. The right side of the device (see area enclosed by white dashed lines) is partially lifted which allows for the ingress of solvents underneath the stack which is seen by the interference pattern. (c) Optical micrograph of a different device after removing the membrane. The hBN (green color) largely covers the electrodes and the graphene (position marked with dashed lines).}
	\label{fig:fig1}
\end{figure}

In this Letter, we present spin transport studied in bottom-up fabricated graphene devices that pushes the measured room temperature spin lifetimes in single layer graphene by a factor of 6 from around $\unit[2]{ns}$ to $\unit[12.6]{ns}$ yielding spin diffusion lengths of $\unit[30.5]{\mu m}$. These values exceeds current models for contact-induced spin dephasing ruling out  spin absorption as the the dominant source of spin dephasing. The major improvement in the device fabrication was triggered by the observation that only a large hBN flake with the attached graphene flake far away from its edges can protect graphene from solvents which are needed to dissolve the supporting PMMA membrane after dry transfer of the PMMA-hBN-graphene stack (see illustration of a finished device after removing the PMMA membrane in Figure \ref{fig:fig1}a). When using smaller hBN flakes, we typically observe a partial lifting of the hBN flake at the side of the device which favors the ingress of solvents along the substrate-to-hBN interface (seen as interference patterns in the optical image in Figure \ref{fig:fig1}b). We find that these devices exhibit spin lifetimes less than $\unit[1]{ns}$ which we attribute to contaminations and deteriorations of the graphene/SiO$_2$ and graphene/MgO/Co interfaces by the solvents. On the other hand, the key device discussed in this paper shows spin lifetimes up to $\unit[12.6]{ns}$. It was built with a very large hBN flake allowing for better protection of the graphene flake (see optical micrograph of this device in Figure \ref{fig:fig1}c). Surprisingly, these long spin lifetimes were achieved despite the existence of conducting pinholes throughout the spin injection and detection MgO barriers as imaged by conductive scanning force microscopy. We show that the magnitude of the contact-resistance-area product ($R_\text{c}A$) does not determine the measured spin lifetimes in all bottom-up devices. Moreover, a recent model including contact-induced dephasing by spin absorption\cite{PhysRevB.91.241407} cannot describe our data suggesting that intrinsic graphene properties get more important in our best devices.

The sample fabrication consists of two main steps. In a first step we use standard electron beam lithography, molecular beam epitaxy (MBE) and a lift-off process to define the electrodes on a Si/SiO$_2$ (285~nm) chip. We use $\unit[35]{nm}$ Co followed by $\unit[1.2]{nm}$ of MgO. The latter is used as a spin injection and detection barrier. In a second step, we exfoliate graphene from natural graphite onto a second Si/SiO$_2$ chip which is dry-transferred on top of these predefined electrodes by an exfoliated hBN flake supported by a PMMA membrane. In a last step the PMMA membrane is dissolved by submerging the device in acetone and isopropanol. When using a flat beaker and an optical microscope with a long working distance, the diffusion of solvents towards the graphene flake can directly be seen by the appearance of an interference pattern in the optical image (see area within the white dashed lines in Figure \ref{fig:fig1}b). Depending on the size of the hBN flake and its adhesion to the underlying substrate this ingress of solvents may or may not occur. We note that after drying the sample stack with nitrogen it is no longer possible to optically distinguish if the hBN flake was previously lifted or not. We also note that the hBN flake is generally not fully adapting to the sidewalls of the electrodes or is even suspended between neighboring electrodes if their separation is small enough (see illustration in Figure \ref{fig:fig1}a).\cite{Droegeler2014} As a result there will be cavities between the substrate and the hBN flake which may promote the diffusion of solvents and residues of PMMA from the edges of the hBN flake along the cavities towards the graphene flake. We find that large hBN flakes which substantially cover the whole electrode structure significantly hinder diffusion of solvents along these cavities.

\begin{figure}[t]
	\includegraphics*[width=\linewidth]{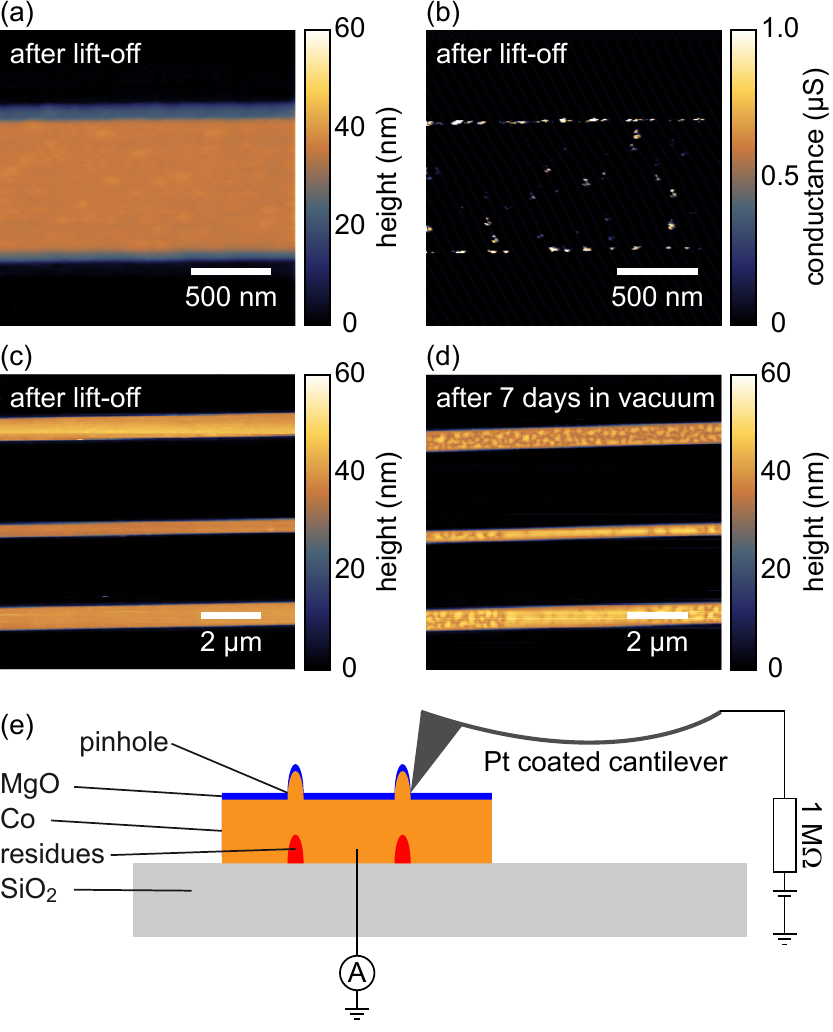}
	\caption{(Color online) (a) Topography and (b) conductance of one of the MgO/Co electrodes after lift-off. The maximum conductance of $\unit[1]{\mu S}$ is determined by a $\unit[1]{M\Omega}$ resistor which is put in series in the current path in order to protect the CSFM tip from high currents. (c) Topography of the electrodes right after lift-off and (d) after storing the sample for seven days in vacuum ($\unit[10^{-4}]{mbar}$). (e) Schematic illustration of the measurement setup. Additionally, the effect of directional growth of the MgO layer on the measured conductance is depicted.}
	\label{fig:fig2}
\end{figure}

\begin{figure}[t]
\centering	
\includegraphics*[width=\linewidth]{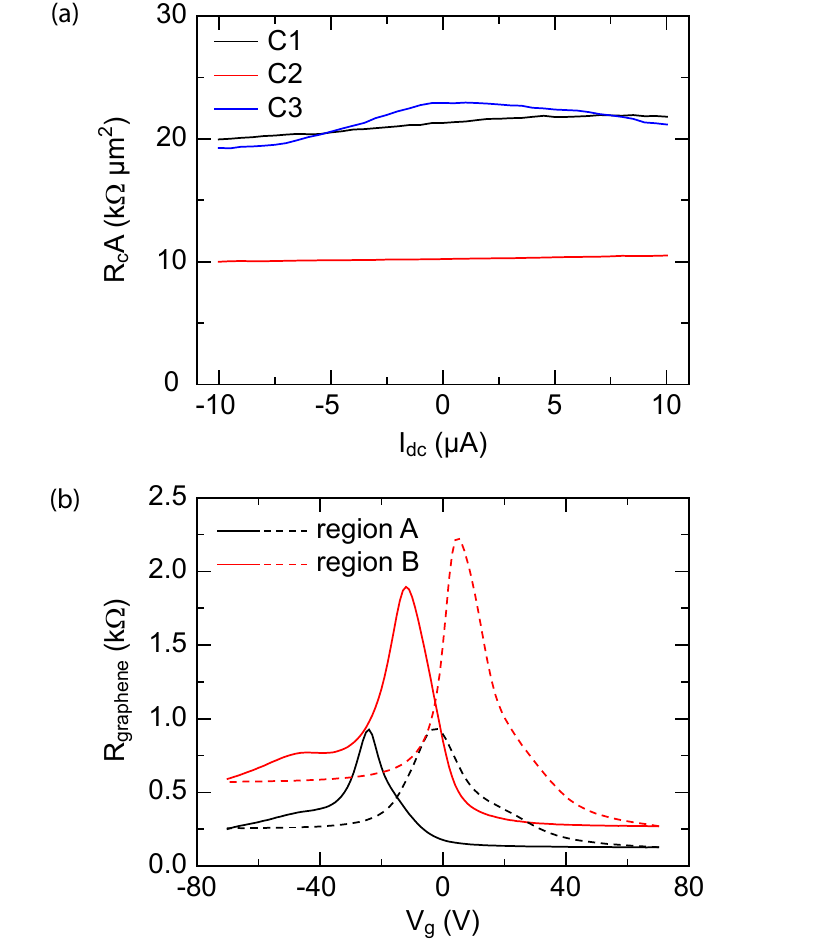}
	\caption{(Color online) (a) $R_\text{c}A=\text{d}V/\text{d}I\cdot A$ curves of the contacts used for spin injection and detection in regions A and B in Figure \ref{fig:fig1}c. (b) Gate dependent graphene resistance for region A (black curves) and B (red curves) where the trace (solid lines) is swept from $V_\text{g}=\unit[-70]{V}$ to $\unit[70]{V}$ and the retrace (dashed lines) is swept from $V_\text{g}=\unit[70]{V}$ to $\unit[-70]{V}$. }
	\label{fig:fig3}
\end{figure}

First, we focus on the electronic properties of the MgO barrier by spatially mapping its  surface right after the lift-off process using conductive scanning force microscopy (CSFM). We show that the MgO barrier exhibit conducting pinholes which result in a flat $\text{d}V/\text{d}I$ characteristic of the electrodes (see Figure \ref{fig:fig3}a). As a measurement tip we use a platinum coated silicon cantilever in contact mode. We apply a voltage of $\unit[200]{mV}$ between the tip and the electrodes which are bonded to a chip carrier and are connected to a $\unit[1]{M\Omega}$ resistor and a current amplifier in series to measures the current to ground (see Figure \ref{fig:fig2}e). The $\unit[1]{M\Omega}$ resistor limits the current in order to protect the metal coating of the cantilever. Hence, the maximum conductance of $\unit[1]{\mu S}$ corresponds to a metallic contact. The topography of one electrode is shown in Figure \ref{fig:fig2}a. The corresponding conductance map is depicted in Figure \ref{fig:fig2}b. It shows randomly distributed hot spots of high conductance which we attribute to pinholes within the MgO layer. Furthermore, the edges of the MgO/Co electrode appear to have a large density of conductive hot spots, but since the MgO layer only covers the top of the Co layer it is likely that the cantilever touches the side of the Co that is not covered by MgO when scanning across or along its edges (see Figure \ref{fig:fig2}e). Some pinholes can be linked to slight local changes in the electrode thickness of $\unit[1-2]{nm}$ which is typically larger than the MgO layer thickness of 1.2~nm. These local elevations may be caused by residues on the substrate prior to metallization. Due to the directional MBE growth, the MgO layer cannot fully cover the edges of these elevations which explains the existence of high conductive local pinholes with potentially direct contact of the cantilever to the underlying Co (see Figure \ref{fig:fig2}e). It has been suggested that these pinholes reduce the spin injection and detection efficiency and may lead to additional spin scattering.\cite{PhysRevB.90.165403}. We note that the remaining parts of the MgO surface is not conducting within the resolution of our setup. As a result, the measured contact resistance between the electrodes and graphene in the final device will be dominated by the conducting pinholes.

We next explore the long term stability of the electrodes by storing the sample under moderate vacuum conditions ($\unit[10^{-4}]{mbar}$) for seven days. For all electrodes we observe a strong change in the surface morphology (compare Figs. \ref{fig:fig2}c and \ref{fig:fig2}d). The MgO surface layer clusters and the average total thickness of the electrodes increase by a couple of nm. In contrast, we do not observe such clustering on a pure Co reference sample (not shown). Since MgO is known to be hygroscopic, we attribute its surface clustering to a reaction of MgO with residual water in the vacuum chamber forming magnesium hydroxide.\cite{Liu1998} For the spin transport devices, it is therefore of utmost importance to transfer the hBN-graphene stack as fast as possible after metallization to prevent MgO from forming magnesium hydroxide which strongly increases the effective thickness of the MgO barrier. We note that the electrodes shown in Figure \ref{fig:fig2}d do not exhibit any conductance in CSFM measurements (not shown). Remarkably, all finished spin transport devices only exhibit small changes of the contact resistances over time even when stored under ambient conditions demonstrating that the graphene-hBN stack protects the MgO layer from forming magnesium hydroxide.

We now discuss charge and spin transport measurements on the device shown in Figure \ref{fig:fig1}c which did not show any ingress of solvents when dissolving the PMMA membrane. All measurements were carried out under vacuum conditions at room temperature using standard low frequency lock-in techniques.\cite{PhysRevB.88.161405} We explore two regions with a transport length of $\unit[3]{\mu m}$ for region A and $\unit[6.5]{\mu m}$ for region B as depicted in Figure \ref{fig:fig1}c. We note that region A is part of region B. Hence, in region B the spins have to pass the floating center electrode C2 when traveling from the spin injection contact C1 to the spin detection contact C3. Comparing the spin transport properties between both regions thus allows to probe additional spin scattering caused by contact C2 when the spin current diffuses along its Co/MgO-to-graphene interface.

In Figure \ref{fig:fig3}a we show the differential contact-resistance-area products $R_\text{c}A=\text{d}V/\text{d}I\cdot A$ of contacts C1 to C3 with $A$ being the respective contact areas. For this measurement we use a three-terminal configuration as described in the supplement of Ref. \citenum{PhysRevB.88.161405}. The overall $R_\text{c}A$ value at $I_\text{dc}=\unit[0]{\mu A}$ is comparable to previous devices fabricated by our bottom-up approach. In agreement with the CSFM measurements, the $R_\text{c}A$ values are almost independent of the applied dc bias indicating ohmic contacts via the conducting pinholes rather than tunneling transport through the MgO barrier which should result in a pronounced increases of $R_\text{c}A$ towards zero bias.
\begin{figure*}[t]
\centering	
\includegraphics*[width=\linewidth]{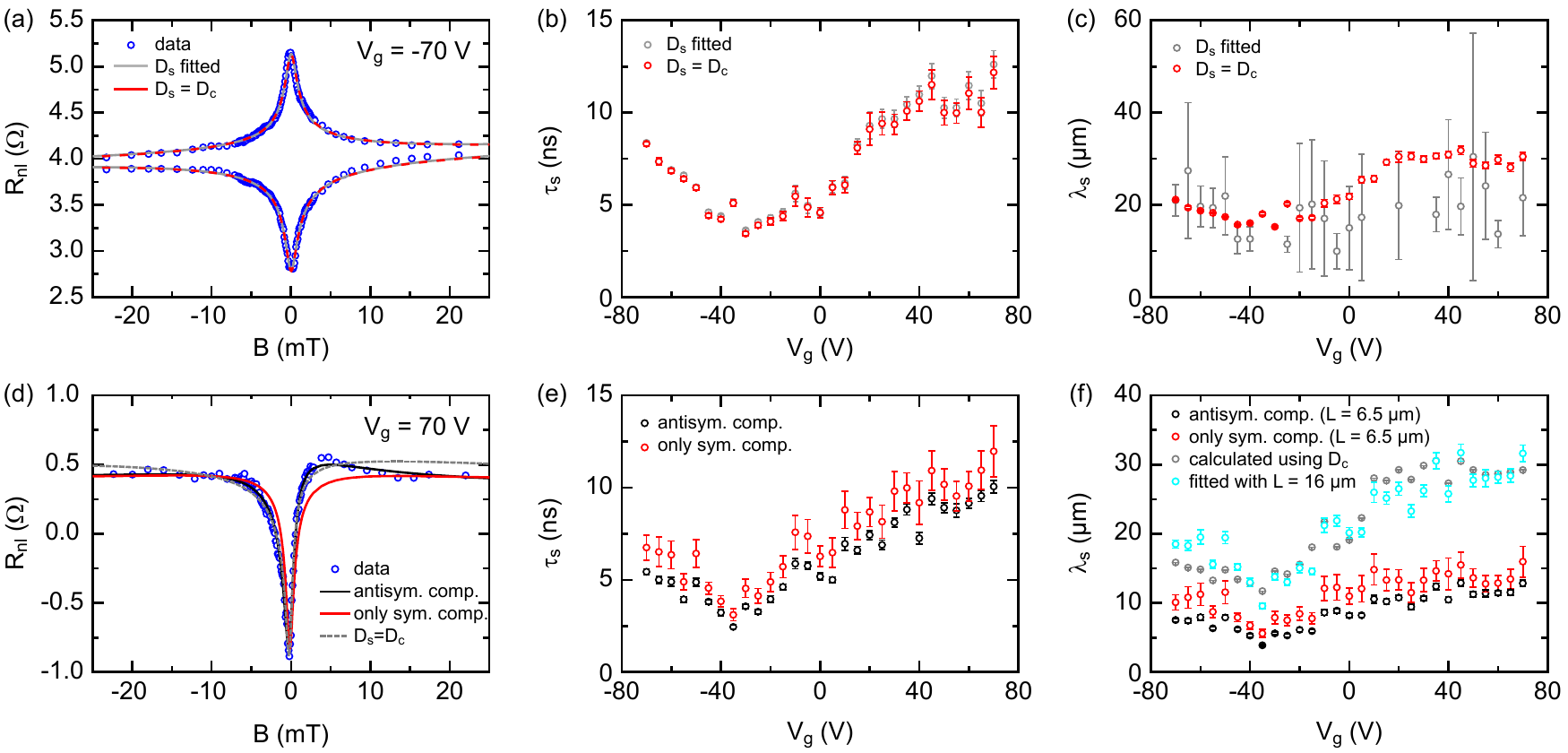}
	\caption{(Color online) (a) Room temperature Hanle spin precession curves measured in region A at $V_\text{g} = \unit[-70]{V}$. The circles represent the data points whereas the dashed and solid lines corresponds to the standard Hanle fit with $D_\text{s}$ being a free parameter (gray line) and $D_\text{s}=D_\text{c}$ (red line) which we extract from the graphene resistance in Figure \ref{fig:fig3}b. (b) Gate dependence of $\tau_\text{s}$ for both fits shown in (a). $\lambda_\text{s}$ vs. $V_\text{g}$ from fitting $D_\text{s}$ (grey circles) and from using $D_\text{s} = D_\text{c}$ (red circles). (d) Hanle spin precession curve for region B at $V_\text{g} = \unit[70]{V}$ for antiparallel alignments of injector and detector electrode magnetization. The red line shows the usual symmetric Hanle fit whereas the black line additionally includes an antisymmetric contribution to the Hanle curve. The dashed gray line results from a fit with $D_\text{s}=D_\text{c}$ (see text). (e) and (f) show the gate dependence of $\tau_\text{s}$ and $\lambda_\text{s}$, respectively.  Panel f additional includes fits for a different transport channel length of $L=\unit[16]{\mu m}$ (cyan circles) and for setting $D_\text{s}=D_\text{c}$ (gray circles).}
	\label{fig:fig4}
\end{figure*}

The room temperature four terminal graphene resistance versus backgate voltage $V_\text{g}$ is shown in Figure \ref{fig:fig3}b. For both regions there is a significant hysteresis between the trace (solid line) and retrace (dashed line) measurements. For the trace scans there are additional charge neutrality points (CNP) at large negative gate voltages. These features can be assigned to trapped charges which are charged and discharged by the backgate voltage.\cite{10.1063/1.3626854} They are only seen if the gate voltage is swept beyond $\unit[\pm30]{V}$. In contrast, traces for smaller gate voltages do not show any hysteresis and exhibit only one CNP close to zero gate voltage. Since graphene on SiO$_2$ usually does not show such a feature we believe that it is caused by polymer residues from the electrode fabrication or trapped water molecules on the SiO$_2$ surface after lift-off.

The charge carrier mobility $\mu$ is extracted from the gate dependent conductance $\sigma$ by $\mu = 1/e \cdot \Delta \sigma / \Delta n$ where $e$ is the electron charge. The charge carrier density $n$ is given by $n=\alpha(V_\text{g}-V_\text{CNP})$, where $V_\text{CNP}$ is the gate voltage at the CNP and $\alpha$ the capacitive coupling to the backgate. We note that the electrodes partially shield the electric field of the backgate. Therefore, we calculated an effective $\alpha = \unit{4.8}\cdot \unit{10}^{10}~ \text{V}^{-1} \unit{cm}^{-2}$ from an electrostatic simulation of our structure. Because of this shielding, the chemical potential of the graphene parts which are on top of the electrodes cannot be tuned by the back gate. These graphene parts contribute a gate-independent resistance to the overall measured graphene resistance.\cite{Volmer2015Synthetic} Both, the gate-independent resistance and the additional CNPs result in a smaller slope of $\sigma$ versus $V_\text{G}$. The extracted mobilities are thus conservative lower bounds of the actual values. For the curves shown in Figure \ref{fig:fig3}b we extract mobilities of $\unit[18,000]{cm^2/(Vs)}$ and $\unit[21,000]{cm^2/(Vs)}$ for regions A and B, respectively.

\begin{figure}[t]
\centering	
\includegraphics*[width=\linewidth]{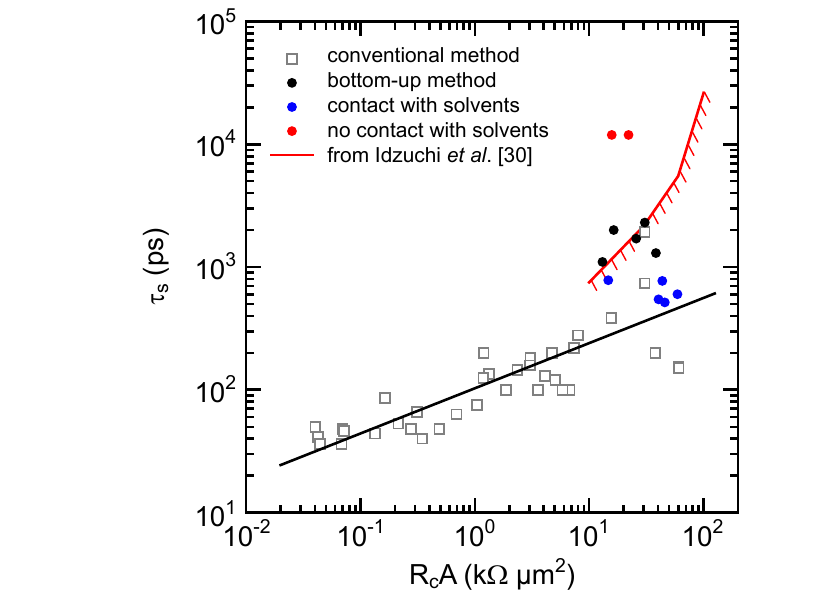}
	\caption{(Color online)  Spin lifetime versus contact-resistance-area product of respective injection and detection electrodes at room temperature. For comparison we included all previous results from Refs. \citenum{PhysRevLett.107.047206} and \citenum{PhysRevB.88.161405} on single layer graphene and bilayer graphene devices that were fabricated by the conventional method where the electrode are directly deposited onto the graphene layer- (squares). The solid line is a guide to the eye. Previous results from single layer graphene bottom-up devices are taken from \citenum{Droegeler2014} (black dots). For devices with the blue dots the graphene got into contact with solvents when dissolving the hBN-graphene supporting PMMA membrane while for the devices with the longest spin lifetime of 12.6ns (red dots) the graphene was fully protected by a large hBN flake. The red solid line is adapted from Ref. \citenum{PhysRevB.91.241407}. It illustrates the longest measurable spin lifetimes if spin absorption by the contacts is taken into account.}
	\label{fig:fig5}
\end{figure}

For spin transport measurements, we use the standard four terminal non-local measurement scheme as described in Ref. \citenum{PhysRevB.88.161405}. To extract the spin lifetime $\tau_\text{s}$ as well as the spin diffusion coefficient $D_\text{s}$ we perform Hanle spin precession measurements where an external magnetic field is applied perpendicular to the graphene flake. The spin diffusion lengths $\lambda_\text{s}$ are calculated by $\lambda_\text{s}=\sqrt{\tau_\text{s} D_\text{s}}$. A typical room temperature Hanle spin precession measurement is shown in Figure \ref{fig:fig4}a for region A at $V_\text{g} = \unit[-70]{V}$ for both parallel and antiparallel alignments of the respective magnetization direction of the ferromagnetic injector and detector electrodes. For fitting the data, we use a simplified solution of the steady state Bloch-Torrey equation with an additional parabolic background to the data (expression in Ref. \citenum{PhysRevB.90.165403}). The background is mainly caused by inhomogeneous injection and detection due to pinholes.\cite{2053-1583-2-2-024001} In our data analysis, we follow a conservative approach to fit the Hanle curves by neglecting corrections to the fit formula which consider contact-induced spin scattering effects, in particular spin absorption. As a result, the extracted spin lifetimes are  underestimated\cite{PhysRevB.86.235408,PhysRevB.89.245436,PhysRevB.91.241407,2015arXiv151202255A} and hence all values given in this work should be considered as lower bounds.
In this context, the extracted spin lifetime of $\tau_\text{s}=\unit[7.7]{ns}$ and spin diffusion length of $\lambda_\text{s}=\unit[21]{\mu m}$ for the Hanle curve in Figure \ref{fig:fig4}a (fit shown as gray lines) are remarkably long.

Before we discuss the gate dependence of $\tau_\text{s}$ (Figure \ref{fig:fig4}b) and $\lambda_\text{s}$ (Figure \ref{fig:fig4}c) in more details, we notice that there is a large error in the determination of $D_\text{s}$ and therefore $\lambda_\text{s}$ as seen by the grey error bars in Figure \ref{fig:fig4}c. This large errors result from the lack of an overshoot of the non-local resistance which is typically seen at moderate magnetic fields and mainly determines $D_\text{s}$ (see Ref. \citenum{PhysRevLett.107.047206}). We therefore consider the charge diffusion coefficient $D_\text{c}$ from the gate dependent graphene resistance (Figure \ref{fig:fig3}a) using $\sigma = e^2 \nu D_\text{c}$, where $\nu = \frac{\sqrt{g_\text{s}g_\text{v} n}}{\sqrt{\pi} \hbar v_\text{F}}$ is the density of states at the Fermi level with the spin and valley degeneracy of $g_\text{s}=2$ and $g_\text{v}=2$, Planck's constant $\hbar$ and the Fermi velocity $v_\text{F}= \unit[10^6]{m/s}$.\cite{RevModPhys.81.109} As seen in Figure~\ref{fig:fig4}c, we find that $D_\text{s} \approx D_\text{c}$. We therefore refit the Hanle curve in Figure \ref{fig:fig4}a by setting $D_\text{s}=D_\text{c}$ (see red curve) which gives equal fit quality as for the grey curve. The corresponding gate dependent spin lifetimes are plotted in Figure \ref{fig:fig4}b for $D_\text{s}$ as a free fitting parameter (gray data points) and $D_\text{s}=D_\text{c}$ (red data points). It is obvious that both approaches essentially give the same values for $\tau_\text{s}$ which shows the robustness against variations of $D_\text{s}$. The spin lifetime is smallest in the vicinity of the CPN and strongly increases towards larger carrier densities. For $V_\text{g}=\unit[70]{V}$ we even achieve $\tau_\text{s}=\unit[12.2]{ns}$ when setting $D_\text{s}=D_\text{c}$ and $\tau_\text{s}=\unit[12.6]{ns}$ when $D_\text{s}$ is being fitted. These values are a factor of 6 larger than the highest reported room temperature spin lifetimes in graphene and demonstrate the key advancement in our fabrication technique.\cite{Droegeler2014,PhysRevLett.113.086602,Kamalakar2015}. Assuming  $D_\text{s} = D_\text{c}$ we achieve spin diffusion lengths (shown in Figure \ref{fig:fig4}c) of $\unit[15.2]{\mu m}$ at the CNP and $\unit[30.5]{\mu m}$ for $V_\text{g} = \unit[70]{V}$. This is also the largest reported value for graphene-based non-local spin valves.

We next focus on the spin transport in region B where we show the Hanle curve in Figure \ref{fig:fig4}d for the antiparallel alignment of the respective magnetization directions of the spin injection and detection electrodes at $V_\text{g}=\unit[70]{V}$. In contrast to the Hanle curve in region A it is not fully symmetric around $B=\unit[0]{T}$. This is most obvious when comparing the data to the standard symmetric Hanle fit which is represented by the solid red line. This asymmetry has also been observed before by other groups but was mostly neglected.\cite{PhysRevLett.107.047207, PhysRevLett.113.086602} To extract the spin transport parameters, we use a fit model which additionally includes a contribution of an antisymmetric Hanle curve which was calculated in Ref. \citenum{PhysRevB.37.5312}. Both $\tau_\text{s}$ and $\lambda_\text{s}$ values are set to be the same for the symmetric and the antisymmetric contribution. Hence, we add only one free parameter to our fit which describes the magnitude of the antisymmetric contribution. The resulting fit is represented by the black solid line which quite accurately describes the data. These antisymmetric contributions can be caused by domain wall pinning along the electrode which was already observed when using Co electrodes.\cite{Berger2015} As the magnetization reverses throughout the domain wall, it exhibit local magnetization directions which are not collinear with the electrodes axis which may explain the existence of the small antisymmetric contribution to the measured Hanle curve.

The gate dependence of $\tau_\text{s}$ for region B is shown in Figure \ref{fig:fig4}e for both fits. As for region A $\tau_\text{s}$ again exceeds $\unit[10]{ns}$ for large positive voltages. The error on the extracted spin lifetimes when including the antisymmetric component is obviously reduced. Additionally, $\tau_\text{s}$ is slightly overestimated if the antisymmetric contribution is neglected. In both regions we obtain very similar $\tau_\text{s}$ values despite the fact that for region B the injected spin accumulation has to fully diffuse under the floating center contact C2 indicating that this contact has negligible impact on the measured spin lifetimes. The corresponding spin diffusion lengths are shown in Figure \ref{fig:fig4}f. Also here the consideration of the antisymmetric contribution decreases the error and leads to slightly lower $\lambda_\text{s}$ values. In contrast to region A we only obtain a maximum value of $\unit[12.8]{\mu m}$. Surprisingly, for this region $D_\text{c}$ is about one order of magnitude larger than $D_\text{s}$ which can also be seen from the minor fit quality of the Hanle curve in Figure \ref{fig:fig4}d (dashed gray line) when using $D_\text{s}=D_\text{c}$. Previously, such deviations between charge and spin diffusion coefficients were attributed to localized states or magnetic impurities yielding an effective g-factor larger than two.\cite{Maassen2013} In order to match $D_\text{s}$ and $D_\text{c}$ in region B we would need an effective g-factor of around 15. As we found $D_\text{s} \approx D_\text{c}$ in region A and considering similar carrier mobilities in both regions we exclude a significant contribution of localized states or impurities at least for graphene in between the electrodes. For the graphene parts which are covered by the electrodes there might be an exchange interaction across the electrode-to-graphene interface which may lead to different local g-factors. But since the ratio of contact-covered parts and free graphene parts are almost the same for both regions, we also exclude this as an explanation for the different $D_\text{c}$ and $D_\text{s}$ values in region B.

Previously, it was shown that Co/MgO electrodes result in local electron doping of the graphene layer in the contact area which forms lateral pnp-junctions along the graphene channel. Although charge carrier diffusion might be affected by the local electric field along these junction, it is not clear why this should result in different charge and spin diffusion coefficients.\cite{Private-Communication-Fabian} Furthermore, we cannot exclude that the outer reference electrode C4 (see Figure 1c) of the non-local voltage measurement detects a significant amount of the overall spin signal. If this was the case the effective spin transport length would be $\unit[16]{\mu m}$ yielding $D_\text{s}\approx D_\text{c}$ as in region A when assuming $D_\text{c}$ to be homogenous across the whole graphene channel. The resulting spin diffusion length is $\unit[31.6]{\mu m}$ at $V_\text{g}=\unit[70]{V}$ (cyan circles in Figure \ref{fig:fig4}f), which compares well to the results of region A (shown in Figure \ref{fig:fig4}c) and to the calculation of $\lambda_\text{s}$ using $D_\text{c}$ (grey circles in Figure \ref{fig:fig4}f). Importantly, the extracted spin lifetimes are not affected by the assumption of a different transport length.

In Figure \ref{fig:fig5} we compare the measured spin lifetimes from this work (red and blue circles) with previous results from bottom-up fabricated single layer graphene devices (black circles) which are plotted on a log-log scale versus the mean $R_\text{c}A$ value of injector and detector electrodes. For comparison, we additionally include results from a previous series of graphene spin-valve devices where we applied the conventional fabrication method by direct evaporation of Co/MgO onto the graphene surface (grey squares). For the latter devices, the contacts are the bottleneck for spin transport, i.e. they cause contact-induced spin scattering, as seen by the strong dependence of $\tau_\text{s}$ on the  $R_\text{c}A$ product. In contrast, all devices fabricated by the bottom-up fabrication method show a completely different behaviour as their spin lifetime do not seem to depend on their respective $R_\text{c}A$ values. As explained above, the strong increase of $\tau_\text{s}$ to 12.6 ns (red circles in Figure \ref{fig:fig5}) results from the best protection of the graphene-to-MgO interface from solvents during processing which we achieved by using a very large hBN flake. In our previous bottom-up devices (black circles) we typically used much smaller hBN flakes. As we did not optically image those devices when removing the hBN-graphene supporting PMMA membrane, we do not know whether graphene got into contact with solvents. To unambiguously demonstrate the role of solvents on the measured spin lifetime we measured several additional devices with the graphene flake close to the edge of the hBN flake. All those devices exhibit partial lifting and ingress of solvents as shown in Figure 1b. The measured spin lifetimes (blue circles in Figure 5) of those devices are diminished by more than an order of magnitude and vary between 500 and 800 ps showing that the exposure to solvents result in additional spin scattering. Surprisingly, even those devices exhibit similar $R_\text{c}A$ values indicating that the contact resistance is hardly affected by the solvents.

It is interesting to compare our results to a recent theoretical model by Idzuchi et al. for contact-induced spin dephasing by spin absorption.\cite{PhysRevB.91.241407} According to this model there is a theoretical limit of the measurable $\tau_\text{s}$ as a function of $R_\text{c}A$ when using the standard model of analysing Hanle spin precession. We included this limit as a red line in Figure \ref{fig:fig5} and note that several of our devices exceed these values. Hence, spin absorption does not seem to be the dominant source of spin dephasing in our devices. It is important to emphasize, however, that present models assume homogeneous barriers and thus do not include inhomogeneous current flow through the pinholes. We furthermore note that a recent study indicates that even in spin transport devices with short spin lifetime below 1~ns spin absorption may not be the dominant spin relaxation mechanism.\cite{2015arXiv151202255A}

In this context, we would like to point out that room temperature nanosecond spin lifetimes measured by other groups are mainly achieved by increasing the transport channel length which is known to reduce the influence of contact-induced spin relaxation effects.\cite{Kamalakar2015,PhysRevB.92.201410} In our study, however, we use transport lengths as low as $\unit[3]{\mu m}$ which is significantly shorter (factor of 3) compared to other groups \cite{Kamalakar2015,PhysRevB.92.201410} and therefore the influence of contact-induced effects should even be more important in our devices. Despite this fact, we are measuring spin lifetimes of $\unit[12.6]{ns}$ which strongly points to the fact that the impact of the contacts on the measured spin lifetime is in fact highly suppressed in our devices.

In conclusion, we have shown that the measured spin lifetime in bottom-up fabricated single layer graphene non-local spin valves can be significantly increased to 12.6 ns which yields spin diffusion lengths of $\unit[30.5]{\mu m}$ at room temperature. This exceptional device performance was achieved by using large hBN flakes which prevent interfacial ingress of solvents to graphene which was shown to strongly reduce spin lifetimes to values below 1~ns. The long spin lifetimes combined with moderate $R_\text{c}A$ products by far exceed current models of contact-induced spin relaxation and spin dephasing which paves the way towards probing intrinsic spin properties of graphene. Our results are furthermore encouraging for future experiments on spin sensitive high frequency applications as the long spin lifetimes can obviously be achieved at moderate $R_\text{c}A$ values which are needed for impedance matching.

We thank J.Fabian for helpful discussions and S. Kuhlen for help on the figures. We acknowledge funding from the European Union Seventh Framework Programme under grant agreement n${^\circ}$604391 Graphene Flagship.
%


\bibliography{bibliography}
\end{document}